# Anomalous Random Telegraphy Signal in Suspended Graphene with Oxygen Adsorption

*Alexandro de Moraes Nogueira[1]\*, Afsal Kareekunnan[1], Masashi Akabori[1], Hiroshi Mizuta[1,2], Manoharan Muruganathan[1†].*

[1]School of Materials Science, Japan Advanced Institute of Science and Technology, Nomi 923-1211, Japan

[2]School of Electronics and Computer Science, University of Southampton, Highfield, Southampton SO17 1BJ, United Kingdom.

ABSTRACT. Graphene is a promising material for sensing applications because of its large specific surface area and low noise. In many applications, graphene will inevitably be in contact with oxygen since it is the second most abundant gas in the atmosphere. Therefore, it is of interest to understand how this gas affects the sensor properties. In this work, the effect of oxygen on the low-frequency noise of suspended graphene is demonstrated. Devices with suspended graphene nanoribbons with a width ($W$) and length ($L$) of 200 nm were fabricated. The resistance as a function of time was measured in a vacuum and pure oxygen atmosphere through an ac lock-in method. After signal processing with wavelet denoising and analysis, it is demonstrated that oxygen causes random telegraphy signal (RTS) in the millisecond scale, with an average dwell time of 2.9 milliseconds in the high-resistance state, and 2 milliseconds in the low-resistance state. It is also shown that this RTS occurs only at some periods, which indicates that, upon adsorption, the molecules take some time until they find the most energetically favorable adsorption state. Also, a slow-down in the RTS time constants is observed, which infers that less active sites are available as time goes on because of oxygen adsorption. Therefore, it is very important to consider these effects to guarantee high sensitivity and high durability for graphene-based sensors that will be exposed to oxygen during their lifetime.

INTRODUCTION

As a two-dimensional material with semi-metallic properties, graphene has many desirable properties, like high conductivity, high mechanical strength, chemical stability, and large specific surface area.[1] Because of this, graphene is a promising material to be used in applications like biological, mass, pressure, electric field, and gas sensing.[2-9] Even single adsorption/desorption detection has already been demonstrated in suspended graphene.[10,11] Furthermore, graphene devices have low-noise figures, especially in the case of suspended graphene.[12,13]

Since oxygen is the second most abundant gas in the atmosphere, it is expected that graphene devices will be exposed to it in many applications. Also, oxygen monitoring is necessary for environments where combustion is a concern, during pharmaceutical production, brewing, *etc*. Even though in many of these applications, reduced noise is essential for high sensitivity, there aren't many studies of the authors' knowledge that treat the effect of oxygen on graphene noise. Also, it is important to notice that while noise usually is undesirable, it can give valuable information. Noise spectroscopy and random telegraphy signal (RTS) analysis is commonly used to study the oxide interface traps in field-effect transistors,[14] and 1/f noise spectroscopy was already used to study the diffusion of Ne atoms on graphene surface at cryogenic temperatures.[15]

Noise can also be used as a parameter for substance identification and to create single-device electronic noses.[16] Low-frequency noise of graphene on $SiO_2$ substrate can be used to discriminate some organic vapors,[17] and noise in the form of RTS was used to discriminate Protoporphyrin, Zn-Protoporphyrin, and Phosphomolybdic acid in a carbon nanotube.[18] Another possible application is the use of noise levels to improve the response time of resistive graphene gas sensors.[19]

Is important to point out that oxygen acts as a p-dopant in graphene. This doping may be irreversible with time and in presence of moisture, which could be a problem for some applications.[20] It was argued that irreversible doping happens mostly in samples on $SiO_2$ substrate, and that suspended graphene would not suffer from it. However, theoretical work shows that defects can act as sites for oxygen chemisorption, and its effect on suspended graphene is not completely understood.[21,22] Therefore, by deepening the understanding of the effects of oxygen on graphene noise, more information on how the adsorption occurs can be obtained and used to design sensors with higher signal-to-noise ratio and durability.

In this paper, monolayer chemical vapor deposition (CVD) graphene was used to fabricate suspended graphene nanoribbons (GNR) with gold electrodes (Figure 1) on a 300 nm $SiO_2$ substrate. The resistance time series was measured in a vacuum and oxygen environment using an ac lock-in technique[23] (Figure 2). Oxygen adsorption causes an increase in the device noise by random resistance changes in the form of RTS. These random changes may affect the response of devices exposed to $O_2$ and must be taken into consideration in sensors that use graphene.

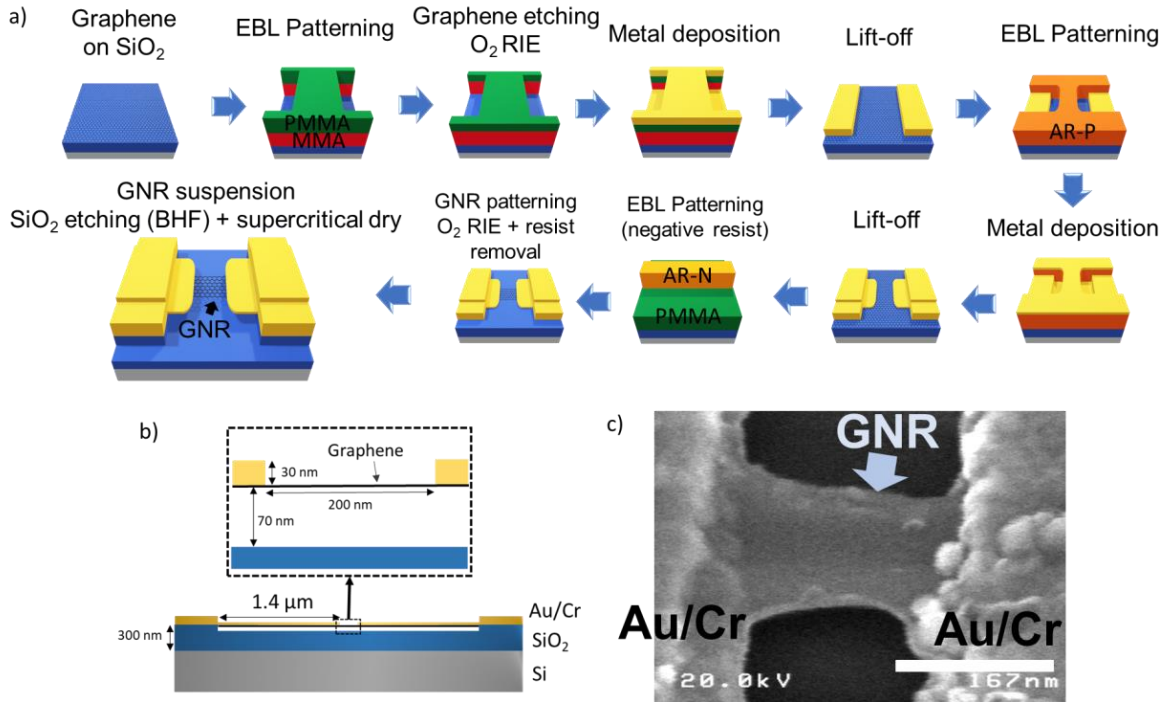

**Figure 1.** (a) Fabrication process schematic (out of scale). (b) Cross-section schematic of the suspended GNR in scale. (c) Scanning electron microscope image of one of the devices showing the suspended GNR in the center and the contacts made of gold and chromium on the sides.

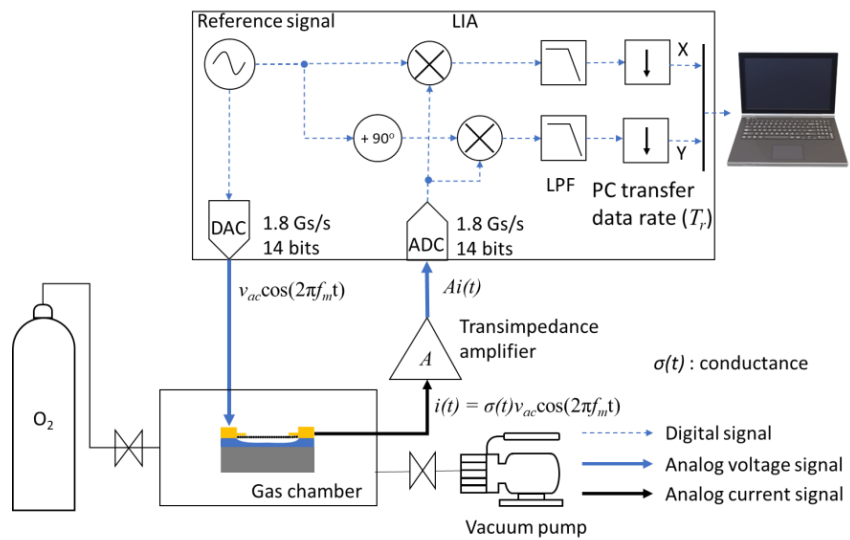

**Figure 2.** The schematic of the setup used to measure the resistance time series in vacuum and oxygen environments.

RESULTS AND DISCUSSION

Suspended GNR devices were fabricated in the same chip. The Fabricated nanoribbons' dimensions are in the nanoscale range because very good gas sensitivity has already been demonstrated in such scales.[8,9] To verify the effect of oxygen, the data of devices that showed RTS in a vacuum and that showed to be collapsed from SEM images were discarded. From the devices measured, two of them (denominated Device A and Device B), showed clear noise distinguishment between the oxygen environment and vacuum. In both devices, the GNR has a width ($W$) and a length ($L$) of 200 nm. Current annealing was performed by applying a drain current of 250 µA in a vacuum of 5 mTorr until the CNP point was around 0 V to clean the devices before the experiments. The drain current ($I_D$) vs gate voltage ($V_G$) curves were obtained by applying a voltage at the bottom of the sample and using the substrate ($SiO_2$ + Si) as the gate. Originally the $SiO_2$ layer had a thickness of 300 nm, but the final thickness is about 230 nm after oxide etching was performed to suspend the graphene. After current annealing, the charge neutral point (CNP) of device A is very close to 0 V, and almost no hysteresis is present in the forward and backward sweeps. This behavior is expected in suspended graphene because there is no effect of oxide interface trap charging. After 5 hours of cooling down in a vacuum, there was a CNP shift in the positive direction (p-doping) so that the CNP is no longer in the measured range.

Similar behavior is known for graphene devices on $SiO_2$ substrate. In that case, the heating of graphene on the $SiO_2$ substrate results in n-doping, and the CNP increases again with cooling in a vacuum.[24] However, the devices in this study are suspended. Thus, there is no effect from the $SiO_2$ and a possible reason for the observed shift is the adsorption of molecules that were adsorbed in regions near the device (metal contacts, $SiO_2$ substrate, *etc.*), and diffused and/or desorbed from their original places to adsorb on the GNR as it was cooling down. Figure 3 (b) shows the $I_D$ vs $V_G$ of device A as it cooled down for 24 minutes, and Figure 3 (c) shows the obtained CNP as a function of time. From Figure 3 (c), it can be seen that the CNP shift is less than 1 V per minute. Since the subsequent time series measurements in a vacuum are also around one minute, the effect of this shift should not be a problem. Also, the shift rate reduces with time, as can be seen by an apparent saturation in the CNP at around 2 V. However, as shown in Figure 3 (a), where the CNP goes outside the measurement range after five hours, the CNP doesn't saturate, its shift only becomes slower.

Before the experiments for Device B, thermal annealing of the entire chip at 150 ºC for 4 hours in a vacuum was performed through a resistive heater. In this case, the entire sample is heated so that molecules adsorbed near the GNR will also be desorbed. However, as can be seen in Figure 3 (d), the CNP is still not in the measured range because the temperature is not enough to desorb molecules that are chemically adsorbed on edges and defect, and current annealing is still necessary. After the current annealing, the CNP enters the measured range and remains there even after 3 hours of cooling down in a vacuum. The same behavior was observed in Device A when thermal annealing was performed (Figure S1). While more investigation on the topic is necessary, it goes beyond the scope of this paper.

Since the devices are suspended, the total gate capacitance ($C_G$) is given by the series arrangement of the capacitance of the $SiO_2$ (230 nm) and the capacitance of the gap between the graphene and the substrate (70 nm). By using a dielectric constant of 3.9 for the $SiO_2$, a total gate capacitance of 6.86 nF/cm$^2$ was obtained. The mobility ($\mu$) can be estimated from the field effect transistor model and is given by $\mu = \frac{\partial I_D}{\partial V_G} \frac{L}{C_G W V_D}$. After annealing and cooling down, both devices presented a transconductance ($\partial I_D / \partial V_G$) of approximately 1 nA/V for $V_G$ lower than -4 V, which gives a mobility of 146 cm$^2$/V.s. It is important to emphasize that the graphene will deflect in the direction of the bottom gate when bias is applied, and the above value is a crude estimative.

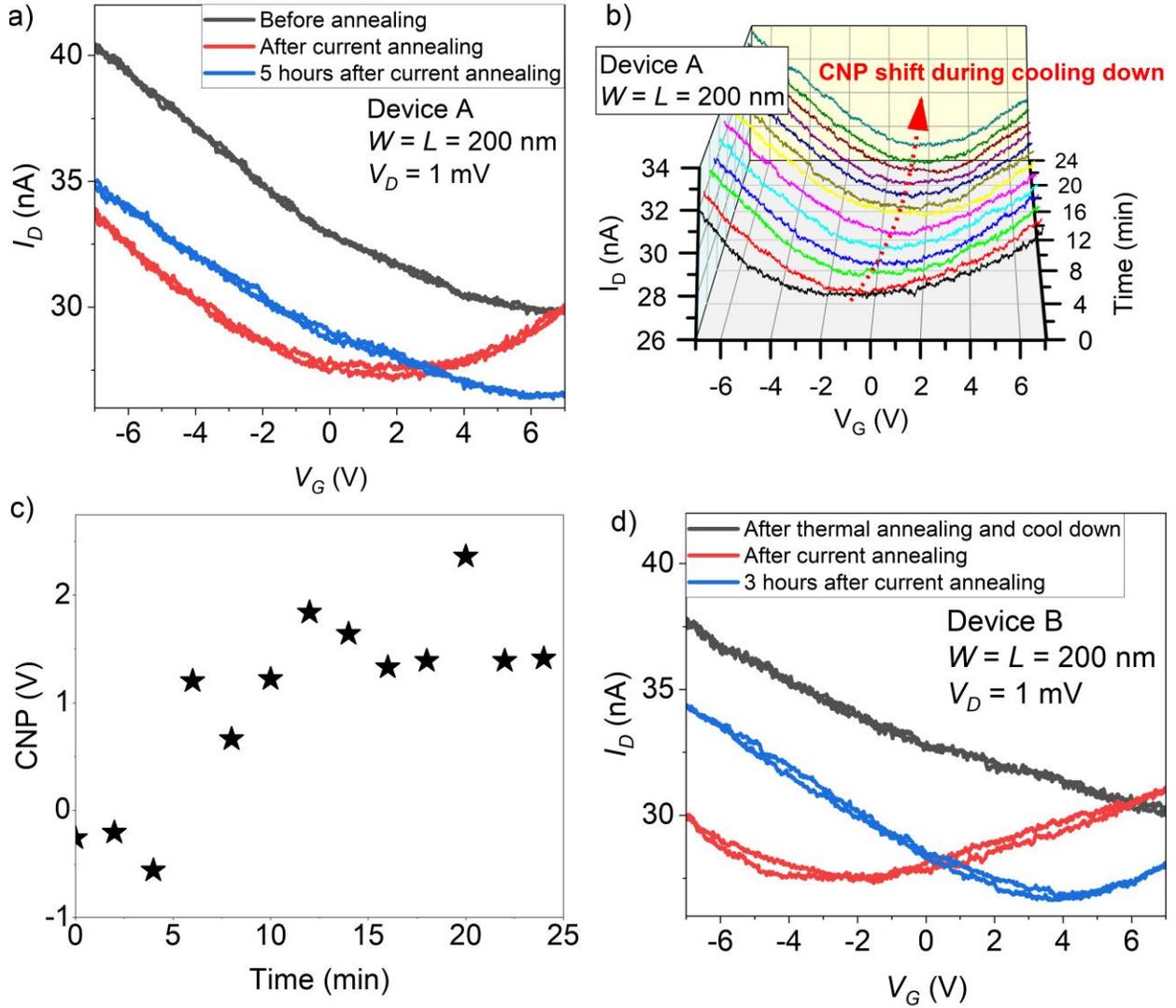

**Figure 3.** Current annealing. (a) $I_D$ vs $V_G$ characteristic of Device A before and after current annealing. (b) $I_D$ vs $V_G$ curves of device A as it was cooling down in a vacuum. (c) CNP of device A as it was cooling down. (c) $I_D$ vs $V_G$ characteristic of Device B before and after current annealing.

After annealing and cooling down for 3 to 5 hours in a vacuum, the resistance measurements with the lock-in amplifier (LIA) were performed by applying an ac signal of 150 mV and frequency of 5 MHz to the drain terminal. No gate voltage was applied. Initially, measurements were taken in a vacuum at a sampling frequency ($F_s$) of 219.7 kHz. For better visualization, Figure 4 (a) shows a sample of 5 seconds of the resistance time series in a vacuum for device A, and Figure 4(b) shows a sample of 5 seconds for device B. In both cases, locally there are no sudden jumps, and the noise seems roughly Gaussian (the entire time series of both devices is shown in Figure S2). After the measurement in a vacuum, oxygen was introduced into the chamber at a rate of 10 sccm. Due to memory concerns, this measurement was performed at a sampling frequency of 429 Hz. Figure 4 (c) shows the resistance time series and the standard deviation calculated in windows of 10 seconds of Device A as oxygen is introduced. As the pressure increases from 6 mTorr to 400 Torr, the noise also increases, especially after 250 seconds (equivalent to a pressure of 370 Torr), where a jump in the standard deviation is visible.

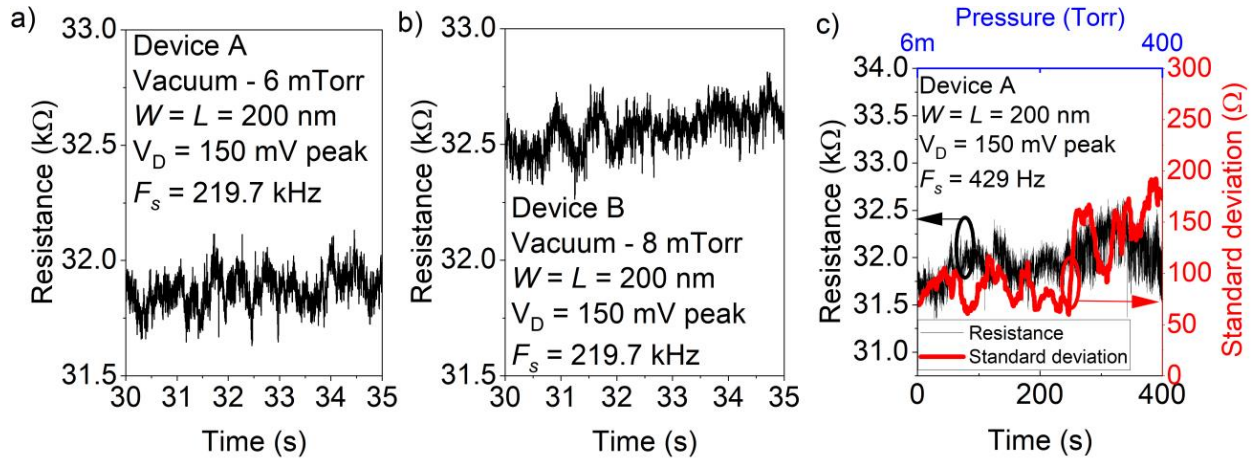

**Figure 4.** The resistance time series in a vacuum for Device A (a) and Device B (b). (c) Resistance time series and the local standard deviation (10 seconds window) as oxygen is introduced into the chamber and the pressure rises from 6 mTorr to 400 Torr for Device A.

The inflow of oxygen was stopped after the pressure arrived at 400 Torr, and new measurements were performed. Figure 5 (a) shows the time series in an oxygen environment for Device A, while Figure 5 (b) shows the time series of Device B. From these measurements it becomes clear that the increase in noise shown in Figure 4 (c) is due to the appearance of RTS, where the resistance changes randomly between a state of higher resistance (up-state) and one with lower resistance (down state). Interestingly, the behavior of the RTS is different for each device. At the beginning of the time series of Device B, the resistance baseline is around 31.2 kΩ most of the time, without large and sudden jumps. However, during some periods denominated as fast RTS, the resistance changes randomly between the original baseline and a lower resistance value of around 30.9 kΩ (a variation of 0.96%). From Figure 5(b), it can also be seen that these fast RTS regions seem to cease at the end of the measurement. For Device A, the transitions are rarer and random at the beginning, but it also shows some of the fast RTS behavior at the end of the time series. This suggests that the RTS in both cases is time dependent and represents the transient behavior of the adsorption of oxygen molecules on the graphene surface.

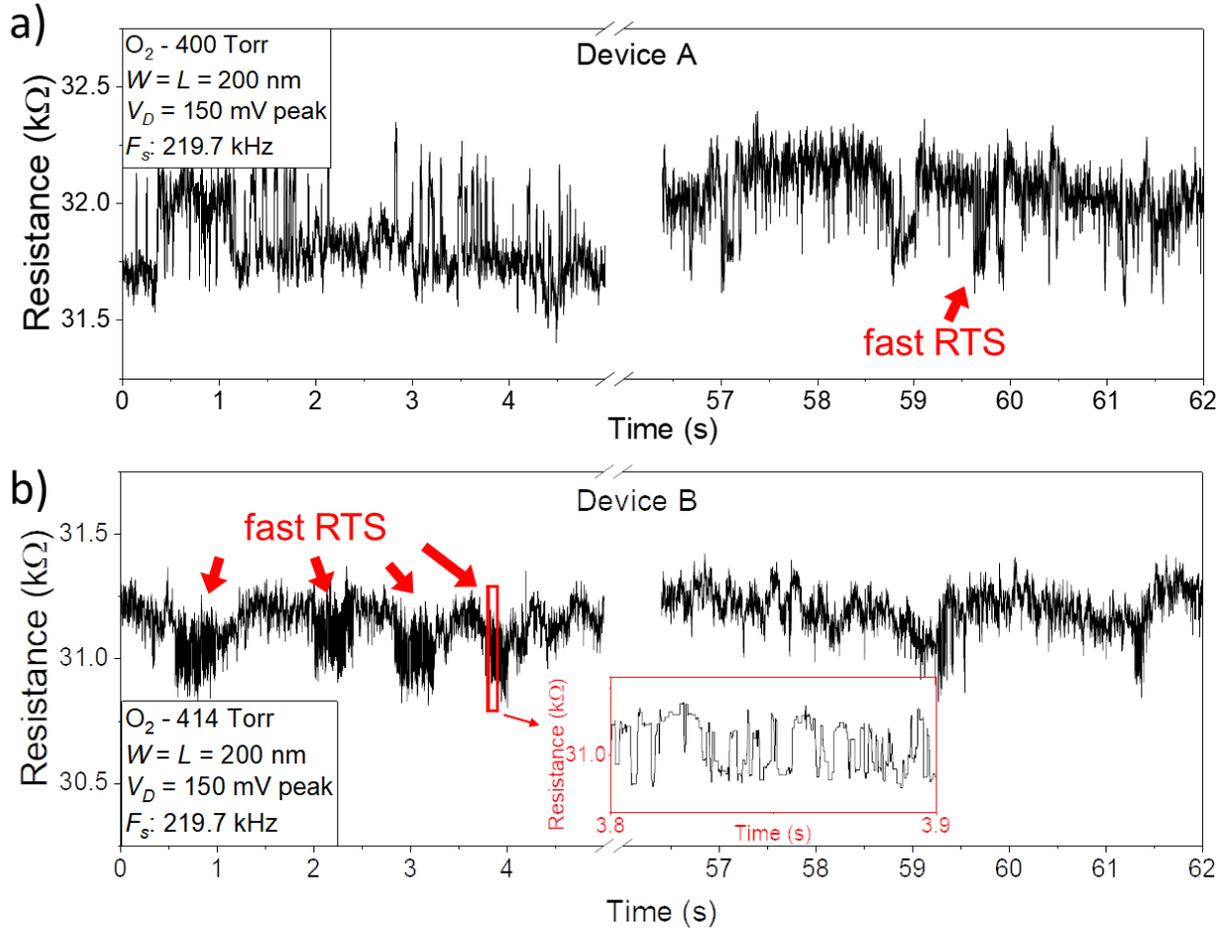

**Figure 5.** Resistance time series in an oxygen environment. (a) Device A Resistance at the beginning and end of the time series. (b) Resistance of Device B at the beginning and end of the time series, showing periods with and without fast RTS. Inset: a portion of the fast RTS.

1/f noise and RTS are usually attributed to mobility variation (Hooge's model[25]) and/or carrier number variations (McWhorter model[26]). In the case of graphene, there is strong evidence that suggests that the 1/f noise is caused by mobility variation.[27,28] RTS with time-dependent behavior like the one observed in Device B is usually denominated as "anomalous RTS" in literature. Anomalous RTS has already been reported for metal-oxide-semiconductor transistors, where it was attributed to oxide charge traps with metastable states.[29,30] However, the analyzed GNRs are suspended; thus, carrier trapping due to defects in the oxide interface cannot explain the measured RTS.

One possible explanation for the RTS is that the adsorption of oxygen creates charge traps. As commented before, device B has a mean resistance value in the low-resistance state of 30.9 kΩ. By using $W = L = 200$ nm and a thickness of 0.3 nm for monolayer graphene[1], the resistivity ($\rho$) was calculated to be 0.927 mΩ.cm. The carrier density ($n$) can be estimated by $\rho = 1/q\mu n$, where $q$ is the electron charge, and the $\mu$ was found previously to be approximately 146 cm$^2$/V.s. A carrier density of 4.63 x 10$^{19}$ cm$^{-3}$ was found for device B. Since the GNR has a volume of 200 nm x 200 nm x 0.3 nm = 1.2 x 10$^{-17}$ cm$^3$, there is a total of 555 charge carriers at one determined moment. If one charge carrier gets trapped, the remaining 554 charge carriers give a charge density of 4.62 x 10$^{19}$, which results in a resistivity of 0.928 mΩ.cm. There is a variation of 0.1% between the high-resistance state and the low-resistance state. This variation is almost one order lower than the observed resistance variation (0.96%). Thus, trapping/detrapping of one charge carrier seems unlikely to be the origin of the RTS.

When an oxygen molecule adsorbs, there is a charge transfer from the graphene to the molecule (0.2 electrons/molecule[31]), along with an increase in Coulomb scattering. More scattering increases the resistance, while p-doping tends to decrease it (for p-doped graphene at $V_G = 0$) such that the two mechanisms will lead to opposite effects. Also, their contribution may depend on the adsorption position, orientation, and the adsorption site itself. Therefore, the anomalous RTS in Device B can be explained if one considers that most of the time (period without RTS), the system with graphene, adsorbate, and gas is in equilibrium until some external disturbance happens. Because of this disturbance, some adsorbed molecules start to oscillate on the graphene. Their distance to the surface changes and the collective contribution of these oscillations causes the variations observed in the total resistance. Since there is the possibility that these oscillations occur in specific defects, it is still not clear if the RTS can be caused by only one molecule or not.

To better analyze the RTS observed in Device B, the regions with and without fast RTS were separated by looking at the envelope of the signal. The reasoning is that where the fast RTS is present, the envelope amplitude (difference between the upper and bottom limits) will be higher. MATLAB function "envelope" was used in the original resistance time series of Device B (Figure 6 (a)), and its amplitude was calculated for each time. The histogram of the envelope amplitude has two distinct peaks (Figure 6 (b)), where the peak at 365 Ω represents the regions where the fast RTS is present. Therefore, by taking only the regions where the envelope amplitude is higher than 300 Ω, the characteristics of the fast RTS can be analyzed. From these regions, the histogram of the resistance deviation from the mean ($\delta R$) was obtained (Figure 6 (c)), and two Gaussian curves were fitted. This fitting can be used in a CUMSUM algorithm,[32] which uses the distribution probability to decide if a transition between two states has occurred (details on the algorithm implementation are present in the Supporting Information, section S6). By knowing when each transition has occurred, the dwell time in the up ($t_{up}$) and down states ($t_{down}$) before each transition can be obtained and stored by the order in which they occur. If the observed RTS is a Poisson process, the probability per unit time of a transition from up to down (down to up) state is given by $1/\tau_{up}$ ($1/\tau_{down}$), where $\tau_{up}$ and $\tau_{down}$ are the means of $t_{up}$, and $t_{down}$, respectively. Also, $t_{up}$ and $t_{down}$ are exponentially distributed and a fitting method can be used to obtain $\tau_{up}$ and $\tau_{down}$.[33] The exponential accumulative function for a variable $x$ and mean $\mu$ has the form $F=1-\exp(-x/\mu)$. By rewriting it as $-\ln(1-F) = x/\mu$, and plotting it as a function of $x$, the slope of the resulting line will be $1/\mu$. Figure 6 (d) shows the probability plot for Device B. The symbols are obtained from the estimated percentiles of the experimental data, and the lines are the respective linear regression.

From the slope of the linear regressions, the following dwell times were obtained: $\tau_{up,} = 2.9$ ms, and $\tau_{down,} = 2$ ms. The asymmetry in the histogram and the dwell times indicate that chemisorption is occurring, such that, under adsorption, the molecules oscillate between two intermediary states (causing the RTS), until they finally find a more stable position and chemisorbs. This would explain why the RTS is only present during some periods.

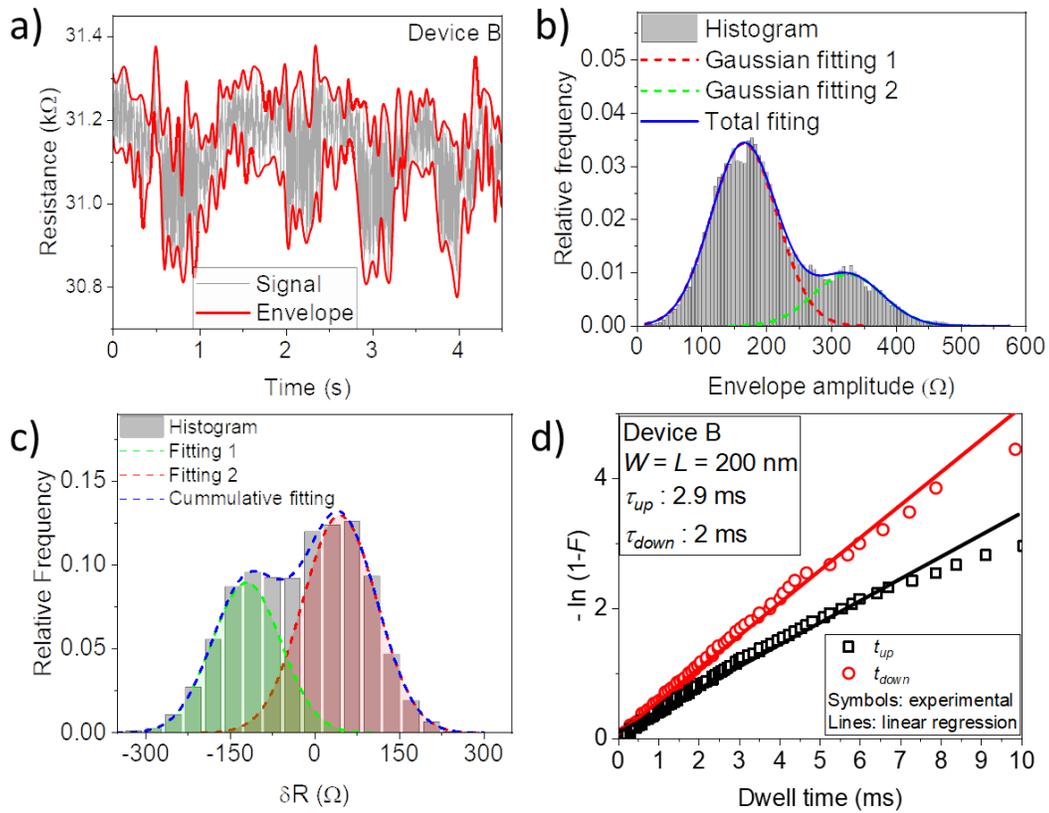

**Figure 6.** RTS analysis of Device B in an oxygen environment. (a) 5 seconds section of resistance-series of the device and its envelope. (b) Histogram of the entire series (62 seconds) of envelope amplitude (difference between the top and bottom of the envelope), where the peak at higher envelope amplitude represents the regions with RTS. (c) Histogram and Gaussians fitting of the resistance deviation ($\delta R$) obtained by considering only periods where RTS is present. (d) Exponential Probability plot of the dwell times in each state obtained from the fast RTS and its linear regression, where $F$ is the cumulative probability.

Since Device A doesn't seem to have periods without RTS, the entire 62 seconds long measurement was analyzed. Figure 7 (a) shows the histogram of the entire time series. The two peaks, showing the up and down states of the resistance are very clear. It also shows that most of the time, the GNR is in the high resistance state, with a higher peak at 32.05 kΩ and a lower one at 31.73 kΩ. This behavior is in accordance with what was demonstrated for Device B. The difference between the two peaks is 320 Ω, a variation of 0.99 % from the high resistance state (very similar to the 0.96 % observed in Device B). All of this suggests that the RTS in both devices has a similar origin. Like in Device B, the CUMSUM algorithm was used to obtain the dwell times.

By putting an index in the transitions, such that the first transition from up to down (down to up) state, gives the first $t_{up}$ ($t_{down}$) at index 1, a plot of the dwell times as functions of the indexes give their temporal behavior. An estimative for $\tau_{up}$ as a function of time was obtained by taking the moving average of the $t_{up}$ vs index number curve with a window of 1 000 points. The left-sided data of Figure 7 (b) (black columns) shows the $t_{up}$, and the right-sided data (red line) shows the estimative for $\tau_{up}$ as a function of the time. Based on that, the $\tau_{up}$ seems to have increased from around 6 ms to approximately 40 ms during the one-minute-long measurement. *i.e.*, transitions from up to down states are becoming rarer with time. Therefore, the RTS has some temporal dependence on both devices. This strongly suggests that the oxygen molecules are chemisorbing. If physisorption was occurring, the molecules could easily desorb at room temperature, and the effect would be independent of time. However, differences in the specifics of each device result in different time parameters (a simple average of the obtained dwell times gives $\tau_{up,}$ = 19.37 ms, and $\tau_{down,}$ = 4.9 ms for Device A).

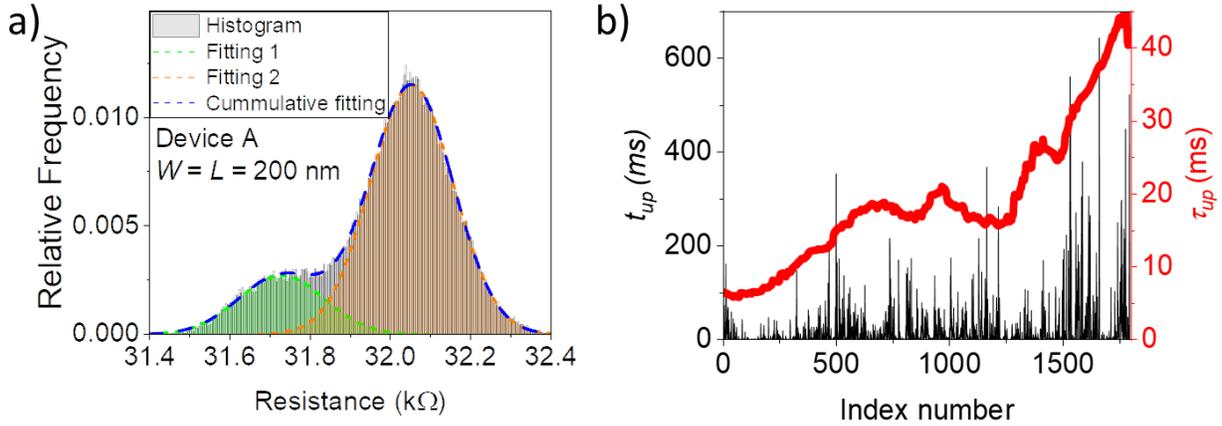

**Figure 7.** RTS analysis of Device A in an oxygen environment. (a) Resistance histogram of the entire 1-minute-long section and fitting with two gaussians. (b) Right-hand (black columns): dwell time in the up state ($t_{up}$) and estimated mean dwell time in the up state ($\tau_{up}$) as a function of the transition index (red line).

A second measurement was performed in Device B one day later, without removing the sample from the chamber. Just after the first experiment, the chamber was evacuated. Thermal annealing was performed at a temperature of 150 °C for 1.5 hours, followed by current annealing. Then, the device was left in a vacuum for 14 hours before the resistance measurements started. Figure 8 (a) shows the $I_D$ vs $V_G$ curves after thermal annealing, soon after current annealing, and 14 hours after the annealing. Just after the current annealing, the CNP became negative (approximately -3V), showing that the current annealing temperature is high enough to desorb chemically adsorbed oxygen. After 14 hours, the CNP is around 5 V, just one volt more than the first annealing sequence in this device, shown in Figure 3 (b). This confirms that the CNP shift after many hours is slow and doesn't interfere with the time series measurement.

Fourteen hours after the current annealing, measurements in a vacuum were performed. The resistance was monitored in a vacuum for 300 seconds at a sampling rate of 429 Hz (Figure S3 (a)). At the end of this monitoring, a 55 seconds-long measurement at a higher sampling rate (219.7 kHz) was performed. Figure 8 (b) shows that sharp transitions with higher resistance and duration of only a few samples are present in a vacuum. The sampling period is approximately 4.5 μs, and it is very likely that these peaks have a shorter duration, but the measurement set-up was able to measure only the longest ones. The chamber used in the experiments has a volume of approximately 2 liters; by using the ideal gas law ($PV = Nk_BT$, where $P$ is the pressure, $V$ is the volume, $N$ is the number of molecules, $k_B$ is the Boltzmann constant and $T$ is the temperature), it is found that at a pressure of 6 mTorr, there are still $3.8 \times 10^{17}$ molecules in the chamber. The transitions to higher resistance could be caused by adsorption/desorption of left-over gas. It is not clear why these sharp transitions aren't present in the first experiment. It is possible that the device became cleaner after more annealing and a longer time in a vacuum, which increased its sensibility.

After the measurements in a vacuum, pure oxygen was introduced into the chamber until a pressure of 400 Torr. As shown in Figure 8 (c), the sharp transitions to a higher resistance that were present in the measurement in a vacuum are no longer present. Since oxygen was introduced and the chamber pressure increased, it is expected that the coverage percentage on graphene has also increased. This coverage may be reducing the effect of fast adsorption/desorption, and the sharp transitions to a higher resistance are not visible anymore. Because of this, the white noise also seems to have been reduced. Figure 8 (c) also shows that RTS became present, though the dwell time in the low-resistance state is much longer (dozens or hundreds of milliseconds) and transitions are rarer than the RTS observed in the first experiment. This is evidence that RTS is caused by oxygen molecules that are adsorbed in specific defects. Annealing will change these defects and the statistical properties of the RTS.

Finally, the chamber was once more evacuated to a vacuum (around 10 mTorr). After 5 minutes a new measurement was performed (Figure 8 (d)). The resistance time series shows a hybrid behavior between the two previous cases. The physically adsorbed molecules can easily desorb once again, and the adsorption/desorption process increases the white noise compared to the measurement in an oxygen environment. Some random transitions to a lower resistance state are still present which is further evidence that these transitions are caused by chemisorbed molecules. For completeness reasons, the entire time series of the three cases are shown in Figure S3.

Figure 9 shows the $I_D$ vs $V_G$ of Device B before the RTS measurements (vacuum), after the measurements in an oxygen environment, and after the device was left in a vacuum of around 5 mTorr for eight hours. In an oxygen environment, a positive shift in the CNP of the Device is visible due to the p-doping of graphene by the gas. If physisorption was occurring, it would be expected that the CNP would shift to lower values after a long period in vacuum, since the molecules would have time to desorb. However, no difference in the $I_D$ vs $V_G$ curve was observed even with the device being left in a vacuum for 8 hours, which confirms the chemisorption of oxygen molecules.

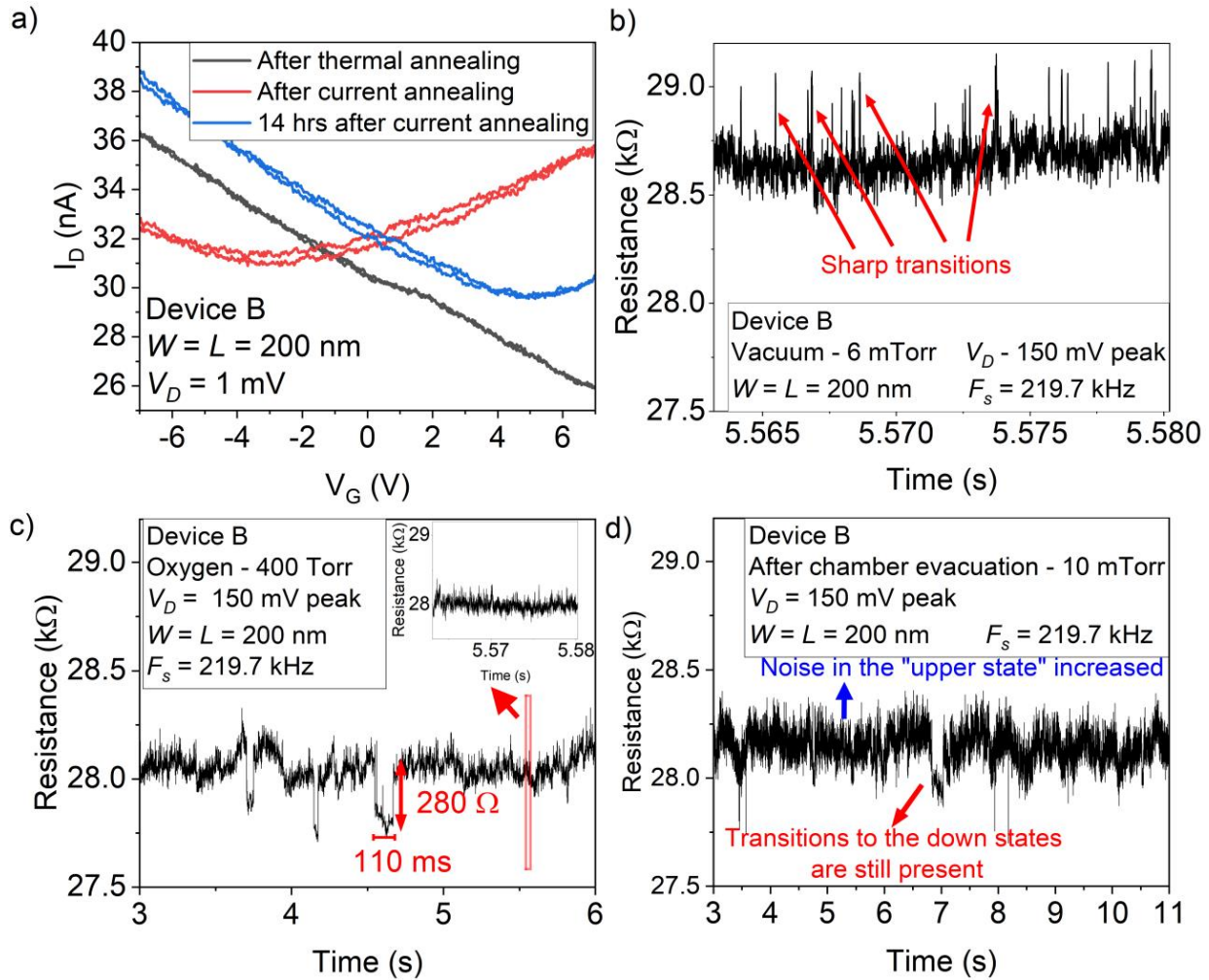

**Figure 8.** Experiment repetition on Device B. (a) $I_D$ vs $V_G$ after thermal and current annealing. (b) Resistance time series segment in a vacuum showing sharp transitions. (c) Resistance time series segment in an oxygen environment showing transitions to lower resistance. Inset: plot at the same scale as the plot in a vacuum, showing the absence of any sharp transitions to a higher resistance state. (d) Resistance time series segment after the chamber was evacuated, where a transition to lower resistance and an increase in the noise if compared to the measurements in an oxygen environment is shown.

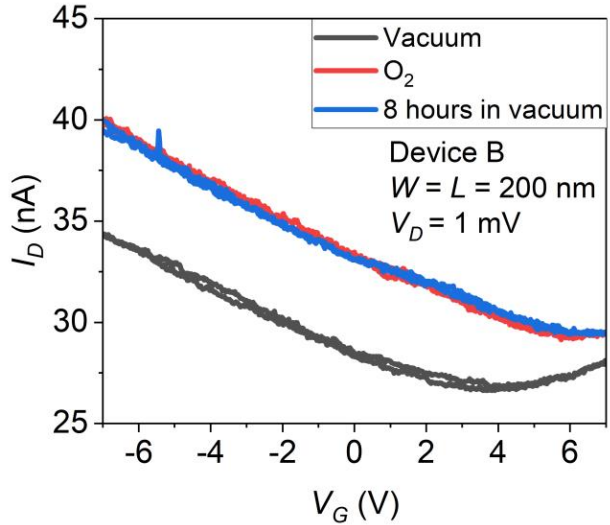

**Figure 9.** $I_D$ vs $V_G$ of Device B. (a) Vacuum environment. (b) Oxygen environment. (c) After 8 hours in a vacuum.

While chemisorption is most likely occurring in defects and edges of graphene, another possibility is that the RTS is caused by chemisorption in the metal contact regions. As can be seen in Figure 1(c), part of the metal contact is also suspended because of the over-etching caused by the graphene beneath. Therefore, there is a possibility that the resistance variation observed is caused by changes in the contact resistance. Also, it was already demonstrated that oxygen chemisorption occurs in graphene defects in a Ruthenium substrate.[34] If that is the case, contact engineering may be used to suppress the RTS, but more investigation is still necessary.

CONCLUSIONS

Anomalous RTS with similar relative resistance variation (0.96% and 0.99%) was observed in suspended graphene nanoribbons in the presence of oxygen for two different devices with dwell times in the order of milliseconds. The origin of the RTS is credited to the transient dynamics of oxygen molecules, where, as they vibrate on the graphene surface their relative distance and effect on GNR resistance will change until they finally chemisorb and the RTS stops. Though it is unclear if this is a collective effect, the effect of very few, or only one molecule. As time goes on, fewer active sites will be available, and a slowdown (less frequent transitions) in the RTS was observed. Chemisorption will change the frequency response and the quality factor of mechanical oscillators, and the random noise may result in false positives in gas sensors. Therefore, these effects cannot be overlooked and need to be considered for applications where the sensor will be exposed to oxygen.

METHODS

**Device Fabrication.** The devices were fabricated from wafers of monolayer graphene on 300 nm $SiO_2$/Si substrate obtained from the company Graphenea. The area for contact pads was patterned using electron beam lithography (EBL) on a bilayer resist of methyl methacrylate (MMA) and polymethyl methacrylate (PMMA). After development, the exposed graphene was etched by oxygen Reactive Ion Etching (RIE), and the metal (5 nm Cr + 75 nm Au) was deposited through electron beam thermal evaporation, followed by a lift-off process in acetone at 60 °C for 30 minutes. Contacts on graphene were patterned with EBL with positive electron beam resist AR-P 6200 for better resolution. 30 nm of metal (5 nm Cr + 25 nm Au) was deposited, followed by a lift-off process with the remover AR 600-71. The graphene nanoribbons (GNR) were patterned through EBL with a bilayer of PMMA and negative resist (AR-N 7520). After development, the exposed PMMA and graphene under it are etched with oxygen RIE, while the negative resist protects the graphene in the nanoribbon area. The remaining resist is removed with acetone and isopropanol rinse. Annealing in an $H_2$ + Ar environment at 300 °C for 3 hours is performed for further cleaning. To suspend the GNR, the sample is rinsed in buffered hydrofluoric acid (BHF) for 45 seconds to etch the SiO$_2$ at an estimated rate of 90 nm/min, resulting in an etching of approximately 70 nm. After dipping the sample in deionized water, it is kept in isopropanol, and supercritical drying is used to avoid the collapse of the GNR.

**Measurement set-up.** Current annealing in a vacuum was performed before the measurements to clean the devices and bring their charge neutral point near 0 V. The $I_D$ vs $V_G$ was measured with the B1500 Semiconductor Device Parameter Analyzer from Keysight at a drain voltage ($V_D$) of 1 mV. The low-frequency resistance fluctuation was measured using the ac lock-in method. An ac voltage signal ($v_{ac}\cos(2\pi f_m)$) at the modulating frequency $f_m$ is generated by the Ultra High-Frequency lock-in amplifier from Zurich instruments and applied in one of the terminals of the device. The resulting ac current ($i(t)$) is the modulation of the applied voltage by the slow variating conductivity ($\sigma(t)$) of the device. A transimpedance amplifier (HCA-400M-5K-C High-Speed Current Amplifier from FEMTO Messtechnik GmbH with gain ($A$) of 5 kV/A) is used to convert the current to voltage so that it can be read by the LIA. After demodulation and filtering by the LIA, one component is obtained by performing the demodulation at the same phase as the reference signal ($X$) and another, with a phase shift of 90º ($Y$). These two components are transferred to a personal computer and the total magnitude of the signal is given by $\sqrt{X^2 + Y^2}$. The current is obtained by dividing the output by the amplifier gain, and the resistance is obtained by dividing the current by the RMS value of the ac input voltage. This method has the advantage of moving the device response to a higher frequency region in which systemic noise is less pronounced. In all the time series measurements, an ac signal of 150 mV peak and frequency of 5 MHz was applied to the drain. No voltage was applied to the bottom gate. The bandwidth (BW) of the low-pass filter (LPF) and the transfer rate ($T_r$) in which the data is transferred to a personal computer were chosen by considering sampling velocity, memory, and aliasing.

**Measurement scheme.** After current annealing in the vacuum and cooling down, measurements in the vacuum were performed with the LIA with $T_r$ =219.7 kHz, and a LPF BW of 78.5 kHz with a duration of around 60 seconds, resulting in approximately $1.3 \times 10^7$ points for each measurement. Next, a new measurement was performed as oxygen was introduced into the chamber at a rate of 100 sccm until the chamber arrived at a pressure of around 400 Torr. Since this measurement takes around 9 minutes, the $T_r$ was reduced to 429 Hz due to memory concerns. The LPF BW was also reduced to 39.18 Hz to avoid aliasing. Finally, a new measurement was performed after the oxygen flow was stopped and the pressure was at around 400 Torr, with the same settings as the one in the vacuum. This entire process was repeated for each device.

**Signal Processing.** The signals in the vacuum and after the flow of oxygen was stopped were processed by applying a Wiener filter and using the *Y* component as an estimative of the measurement system noise (Supporting Information). After the application of the Wiener filter, the noise was further reduced through wavelet denoising. The signal is decomposed in wavelet coefficients, and a threshold is used to remove components with low values. Finally, the signal is recovered by applying the inverse discrete wavelet transform.[35] By doing so, much of the high-frequency noise can be removed, while preserving the sharp transitions that are part of the signal (Figure S3). The envelope of time series shown in Figure 6 (a) was obtained through the MATLAB function "envelope" and the option "peak". In this option, the envelope is obtained from the spline interpolation of maxima and minimum points. The denoising was performed in MATLAB by applying the function "wdenoise". The Daubechies wavelet of level 1 (equivalent to the Haar wavelet) was chosen because of its similarity to a pure RTS. The denoising was performed down to level 19, using a soft universal threshold.

## ASSOCIATED CONTENT

**Supporting Information**. The following file is available free of charge.

Thermal and current annealing of Device A (Section S1), Resistance time series (Section S2), Wiener Filter (Section S3), Wavelet denoising (Section S4), Gaussian fitting of the histograms (Section S5), and CUMSUM algorithm implementation (Section S6) (PDF)

## AUTHOR INFORMATION

**Corresponding Author**

* s2120026@jaist.ac.jp

**Present Addresses**

† TAIYO YUDEN CO., LTD

**Author Contributions**

The manuscript was written through the contributions of all authors. All authors have approved the final version of the manuscript.

**Notes**

The authors declare no competing financial interest.

## ACKNOWLEDGMENT

This work was partially supported by the Japan Ministry of Education, Culture, Sports, Science and Technology (MEXT) scholarship (200469) and by Grant-in-Aid for Scientific Research KAKENHI 21H01386 and 20K20442 from the Japan Society for the Promotion of Science

## ABBREVIATIONS

RTS, random telegraphy signal; CVD, chemical vapor deposition; LIA, lock-in amplifier; CNP, charge neutrality point; EBL, electron beam lithography; MMA, methyl methacrylate; PMMA, polymethyl methacrylate; RIE, reactive ion etching; GNR, graphene nanoribbon; BHF, buffered hydrofluoric acid; BW, bandwidth; LPF, low-pass filter.